\documentclass[runningheads]{llncs}
 \usepackage[T1]{fontenc}
\usepackage{amsmath}
\usepackage{float}
\usepackage{amssymb}
\usepackage{amsfonts}
\usepackage{enumitem}
\usepackage[misc,geometry]{ifsym}
\usepackage{mwe} 
\usepackage{tabularx}
\usepackage{hyperref}

\usepackage{xcolor}

\usepackage{graphicx}
\usepackage{multirow}
\usepackage{float}
\begin{document}
\title{TSBP: Improving Object Detection in Histology Images via Test-time Self-guided Bounding-box Propagation}
\titlerunning{Test-time Self-guided Bounding-box Propagation}
%

\author{Tingting Yang\and Liang Xiao$^{\textrm{\Letter}}$\and Yizhe Zhang$^{\textrm{\Letter}}$}
\authorrunning{T. Yang et al.}
%
\institute{School of Computer Science and Engineering, Nanjing University of Science and Technology, China\\
\email{xiaoliang@mail.njust.edu.cn, yizhe.zhang.cs@gmail.com}}
\maketitle              
\begin{abstract}
\textcolor{black}{A global threshold (e.g., 0.5) is often applied to determine which bounding boxes should be included in the final results for an object detection task. A higher threshold reduces false positives but may result in missing a significant portion of true positives. A lower threshold can increase detection recall but may also result in more false positives. Because of this, using a preset global threshold (e.g., 0.5) applied to all the bounding box candidates may lead to suboptimal solutions.} In this paper, we propose a Test-time Self-guided Bounding-box Propagation (TSBP) method, leveraging Earth Mover's Distance (EMD) to enhance object detection in histology images. TSBP utilizes bounding boxes with high confidence to influence those with low confidence, leveraging visual similarities between them. This propagation mechanism enables bounding boxes to be selected in a controllable, explainable, and robust manner, which surpasses the effectiveness of using simple thresholds and uncertainty calibration methods. Importantly, TSBP does not necessitate additional labeled samples for model training or parameter estimation, unlike calibration methods. We conduct experiments on gland detection and cell detection tasks in histology images. The results show that our proposed TSBP significantly improves detection outcomes when working in conjunction with state-of-the-art deep learning-based detection networks. Compared to other methods such as uncertainty calibration, TSBP yields more robust and accurate object detection predictions while using no additional labeled samples. \textcolor{black}{The code is available at \url{https://github.com/jwhgdeu/TSBP}}.

\end{abstract}
\section{Introduction}
\label{sec:intro}

\begin{figure}[t]
    \centering
    \includegraphics[width=\textwidth]{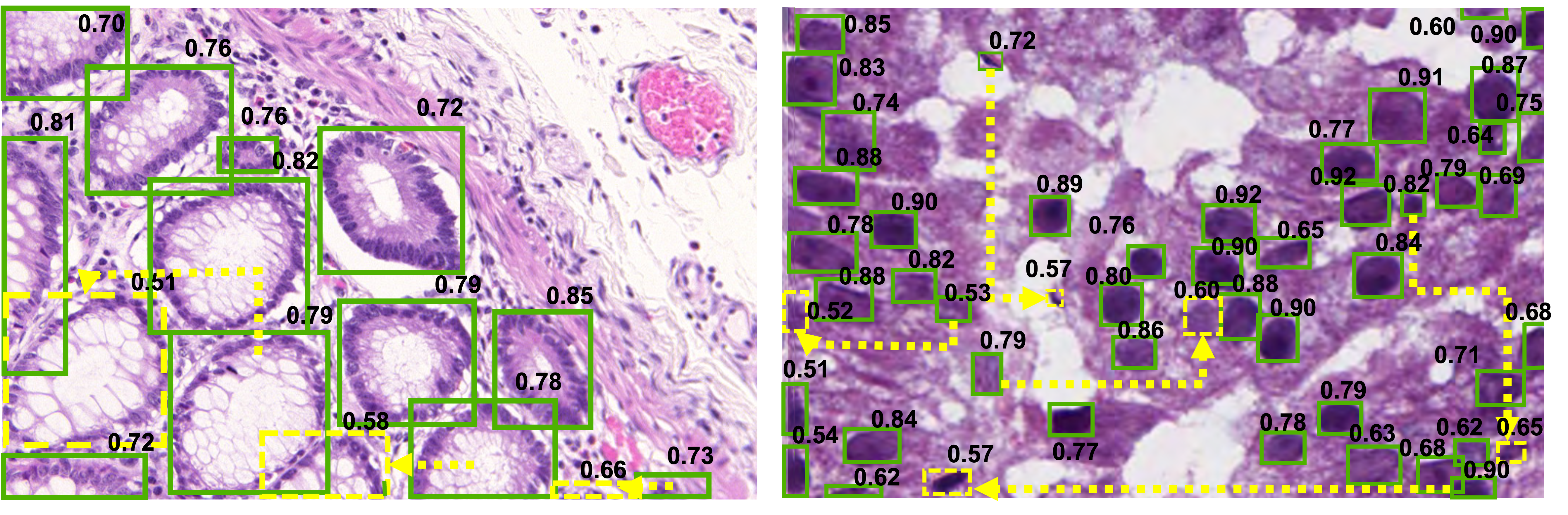}
    \caption{Shown are examples of detected bounding boxes (b-boxes) with varying confidence scores in two test images. The yellow arrows highlight how higher-confident b-boxes influence lower confident b-boxes in TSBP based on visual similarities.}
    \label{fig1}
\end{figure}

Object detection methods enable automated localization and classification of key objects in images, and play an important role in various areas such as medical image analysis and disease diagnosis. In recent years, with the rapid development of deep learning, object detection has made significant improvements in the field of biomedical research, driving innovation in medical image diagnostics and biomedical studies~\cite{davri2022deep,echle2021deep,van2021deep}.

A confidence score of each detected bounding-box is often generated by a detection model, which reflects the level of certainty that the detection is correct\textcolor{black}{~\cite{SODUD}}. Fig.~\ref{fig1} showcases the predicted results of object detection models on two histology images. Higher confidence thresholds (e.g., 0.90) can effectively reduce false positives but lead to false negatives, while lower thresholds (e.g., 0.6) can increase the recall rate at the expense of more false positives. During the deployment (test-time) of an object detection model, using confidence thresholds to determine which bounding-boxes to include is critical but fewer studies were conducted in this aspect\textcolor{black}{~\cite{Confidence_Score}}. A common approach involves testing and evaluating different thresholds on a validation set, followed by selecting the optimal threshold as the deployment confidence threshold~\cite{yue2022deep}. However, this approach has two notable limitations. First, object detection tasks involve diverse objects with varying categories, sizes, and shapes. Consequently, using a single threshold may not effectively cater to all the objects. Second, estimating a threshold using data inevitably incurs the data shift problem\textcolor{black}{~\cite{Domain_Shift}}. That is, when the test set exhibits data shift with respect to the training and validation sets, the estimated threshold can be ineffective for test samples.

Predictions with higher confidence scores are statistically more likely to be correct than those with lower confidence scores. In this paper, we aim to utilize bounding-box predictions with high confidence to rectify the bounding-box predictions with lower confidence. More specifically, we propose a new method called Test-time Self-guided Bounding-box Propagation (TSBP). TSBP propagates information of bounding-boxes with high confidence to influence bounding-boxes with lower confidence, with the goal of adjusting the class labels associated to bounding-boxes with low prediction confidence. Our main contributions can be summarized in the following three aspects.
\vspace{-0.2cm}
\begin{itemize}
\item We explore a practical test-time task aimed at improving object detection performance in histology images. We illustrate the efficacy of a test-time optimization procedure in refining predictions by leveraging visual similarities among test samples. Our findings underscore the potential benefits of this approach in enhancing detection accuracy.

\item To tackle this challenge, we introduce a novel and adaptable method termed Test-time Self-guided Bounding-box Propagation (TSBP). TSBP employs an iterative matching and selection process, leveraging high-confidence bounding-boxes to refine low-confidence ones by exploiting their visual similarities. Unlike conventional confidence calibration techniques, TSBP operates without such requirements for label adjustment. 

\item Experiments, conducted on gland detection and cell detection tasks in histology images, show that our TSBP method can substantially enhance detection outcomes. Compared to known confidence calibration methods, TSBP leads more robust and accurate object detection predictions.
\end{itemize}

\vspace{-0.5cm}
\section{Related Work}\label{sec:related}

Our TSBP method is related to confidence calibration, as TSBP can be viewed as a method for dynamically adjusting (or calibrating) the confidence score for each prediction (bounding-box) of a model. The goal of confidence calibration is to align the predicted confidences of a model with their actual accuracy~\cite{guo2017calibration}. Studies have shown that deep neural networks (DNNs) tend to make overconfident predictions~\cite{deep}, leading to inaccurate confidence estimation. Post-processing calibration, which requires parameter estimation on a validation set, is the most widely-used approach for confidence calibration, and such methods include histogram binning~\cite{HB}, Bayesian binning~\cite{bayes}, Platt scaling~\cite{Prob}, and Beta calibration~\cite{beta}. In~\cite{face}, calibration results were improved by considering both confidence scores and bounding-box information of object detection models. In~\cite{box-Confidence}, histogram binning calibration was modified under the condition of bounding-box sizes.

In traditional classification tasks, the maximum class probability (MCP) is commonly used as the confidence score~\cite{baseline}. This approach assigns the highest softmax output as the confidence score, which means that even incorrect predictions can result in high confidence values. Considering that MCP can lead to the overlap of confidence scores between correct and incorrect predictions, a confidence module called ConfidNet~\cite{addressing} was introduced into the original classification model to output confidence predictions that are closer to the true class probability (TCP). To suppress low-quality prediction boxes, CPSS-FAT~\cite{cpss} multiplies a consistent location quality estimation (CLQE) score by a class score as the confidence score.

\begin{figure}[t]
   \centering
    \includegraphics[width=0.9\textwidth]{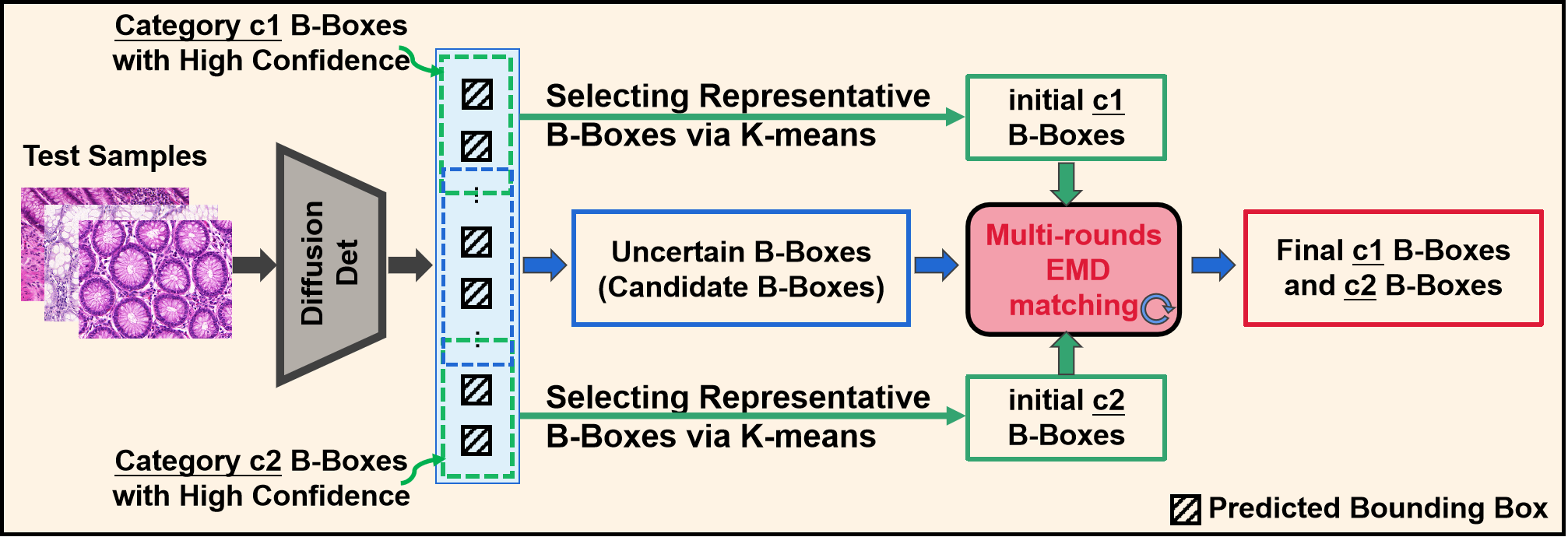}
    \caption{Overview of Test-time Self-guided Bounding-box Propagation (TSBP). }
    \label{fig2}
\end{figure}

\section{Methodology}

Fig.~\ref{fig2} provides a high-level overview of the proposed method. We utilize a trained object detection model (e.g., DiffusionDet~\cite{diffusiondet}) on a set of test images, denoted as $ X = {x_1, x_2, \ldots, x_n}$. For each image $x_i\in \mathbb{R}^{3 \times H \times W}$, we obtain a collection of bounding-boxes $\text{box}_{i,k}={u_{i,k}, v_{i,k}, w_{i,k}, h_{i,k},c_{i,k},s_{i,k}},k=1,2,\ldots,z_{i}$. As the number of generated bounding-boxes may vary for each image during inference, we use $z_i$ to represent the count of bounding-boxes produced for image $x_i$. Here, $(u_{i,k}, v_{i,k}, w_{i,k}, h_{i,k})$ indicates the positional attributes of the corresponding bounding-box: $(u_{i,k}, v_{i,k})$ denotes the top-left corner coordinates, while $(w_{i,k}, h_{i,k})$ represent its width and height, respectively. $c_{i,k}$ denotes the category label of the bounding-box, while $s_{i,k}$ represents its confidence score, ranging from 0 to 1. A high value of $s_{i,k}$ indicates the model's strong confidence in the category label assignment, whereas a low value suggests being uncertain. Below, we describe our proposed method for refining the label assignment of the bounding-boxes, utilizing the higher-confidence bounding-boxes to influence those with lower confidence.

\subsection{Initialization}
\label{sec:Initialization}
In the subsequent matching process, we need to compare the feature similarity between bounding-boxes. Therefore, we introduce an additional feature information $\text{feat}_{i,k}$ for each bounding-box $\text{box}_{i,k}$. $\text{feat}_{i,k}$ is obtained by cropping the image patch inside the bounding-box $\text{box}_{i,k}$ at the corresponding position on the original image $x_i$, and then applying a feature extractor, e.g., ResNet-50~\cite{resnet-50}, to the cropped image patch. We denote feature extractor as $\pi$, and then the feature information $\text{feat}_{i,k}$ is obtained by $\text{feat}_{i,k} =\pi(x_i[v_{i,k}:(v_{i,k}+h_{i,k}),u_{i,k}:(u_{i,k}+w_{i,k})])$. TSBP takes all the bounding-boxes as its input. The task of TSBP is to reassign class categories to those bounding-boxes with low confidence scores.

For simplicity, suppose there are two class categories, denoted as $c1$ and $c2$. We set confidence thresholds, $s^{c1}$ and $s^{c2}$. For bounding-boxes $\text{box}_{i,k}$ with $c_{i,k} = c1$ and confidence scores $s_{i,k} > s^{c1}$, we consider them as high-confident (HF) $c1$ bounding-boxes. Similarly, for bounding-boxes $\text{box}_{i,k}$ with $c_{i,k} = c2$ and confidence scores $s_{i,k} > s^{c2}$, we consider them as high-confident (HF) $c2$ bounding-boxes. The found HF $c1$ and $c2$ bounding-boxes are set as the initial confirmed $c1$ and $c2$ bounding-boxes. We then select a set of representative bounding-boxes from HF $c1$ and $c2$ bounding-boxes for the following Multi-Round EMD Matching step. \textcolor{black}{To achieve more efficient and effective matching performance, we use} a classic K-means algorithm (the number $K$ of clusters is set as 25 by default). All the bounding-boxes not added to the confirmed bounding-box set are considered as candidate bounding-boxes. 

To ensure the reliability of the subsequent multi-round EMD matching and bounding-box propagation, we proceed to compute distance constraints for each class category. For every HF $c1$ bounding-box, we compute the shortest Euclidean distance between this bounding-box and the other HF $c1$ bounding-boxes. This process is repeated for all HF $c1$ bounding-boxes, and then the average of the found shortest distances is computed to establish the distance constraint for class category $c1$, denoted as $D_{{max}}^{c1}$. Similarly, we execute the same procedure for HF $c2$ bounding-boxes to obtain $D_{{max}}^{c2}$. Both $D_{{max}}^{c1}$ and $D_{{max}}^{c2}$ will be applied in the initial stage of the multi-round EMD matching.

\vspace{-0.1cm}
\subsection{Multi-Round EMD Matching and Bounding-Box Propagation}
The multi-round bounding-box propagation comprises two stages. In the first stage, propagation is carried out with stricter constraints when admitting newly confirmed bounding-boxes (via utilizing the above obtained $D_{{max}}^{c1}$ and $D_{{max}}^{c2}$). This is done to ensure that the bounding-boxes added during this early stage are more likely to have correct label assignments. In the second stage, these constraints are relaxed to allow propagation to occur across a wider range of candidate bounding-boxes.

In each round of propagation, suppose there are $a$ candidate bounding-boxes and $m$ confirmed bounding-boxes. Among these confirmed bounding-boxes, $t$ are of class category $c1$ and the rest, $m - t$, are of class category $c2$. The input to Earth Mover's Distance (EMD) consists of two signatures: $P = {p_1, p_2, \ldots, p_a}$ and $Q = {q_1, q_2, \ldots, q_t, q_{t+1}, \ldots, q_m}$, where $P$ represents the candidate bounding-boxes and $Q$ represents the confirmed bounding-boxes. Specifically, $Q$ comprises $t$ confirmed $c1$ boxes and $m - t$ confirmed $c2$ boxes.

The distance between $p_i$ and $q_j$ is denoted as $D_{i,j}$, reflecting the cost of matching a candidate bounding-box $p_i$ with a confirmed box $q_j$. $D_{i,j}$ is computed using the Euclidean distance between the features of $p_i$ and $q_j$ (features obtained in Section~\ref{sec:Initialization}). The matching flow between $p_i$ and $q_j$ is denoted by $f_{i,j}$, with a potential value of either 0 or 1. The matching optimization goal is:

\begin{align}
    &\min_{f}\sum_{1 \leq i \leq a}\sum_{1 \leq j \leq m}f_{i,j}D_{i,j}, \label{eq:1}
\end{align}
subject to 
\begin{align}
    \sum_{1 \leq i \leq a}f_{i,j} & \leq 1 , \ 1 \leq j \leq m \label{eq:2},\\
    \sum_{1 \leq j \leq m}f_{i,j} & \leq 1 , \ 1 \leq i \leq a  \label{eq:3},\\
    \sum_{1 \leq i \leq a} \sum_{1 \leq j \leq m}& f_{i,j}  = \min(a,m), \\
    f_{i,j} \in \begin{Bmatrix}{0,1} \end{Bmatrix}&, 1 \leq i \leq a , 1 \leq j \leq m \label{eq:5}.
\end{align}

Equation~(\ref{eq:2}) indicates that a confirmed bounding-box ($c1$ or $c2$) can be matched with at most one candidate bounding-box. Equation~(\ref{eq:3}) indicates that a candidate bounding-box can be matched with at most one of the confirmed bounding-boxes. When the matching flow $f_{i,j}$ is 1, it means candidate bounding-box $p_i$ has been successfully matched with confirmed bounding-box $q_j$.

In the \textbf{first stage} of propagation, for each confirmed $c1$ bounding-box associated with $q_j$, where $1 \leq j \leq t$, we identify $p_{i^*}$ such that $f_{i^*,j} = 1$. If $D_{i^*,j}$ is smaller than the distance constraint $D_{max}^{c1}$ (obtained in Section~\ref{sec:Initialization}), we include the corresponding candidate bounding-box $p_{i^*}$ in the confirmed $c1$ bounding-box set. The same procedure is followed for $c2$ bounding-boxes associated with $q_j$, where $t < j \leq m$. The confirmed box set $Q$ is then updated to a new state with the newly added $c1$ and $c2$ bounding-boxes, while the remaining candidate bounding-boxes proceed to the next round of EMD matching (with updated $P$).

The \textbf{second stage} of propagation starts when no new bounding-boxes were added to the confirmed bounding-box sets for both $c1$ and $c2$ categories in the last round of propagation. Consequently, the $D_{max}^{c1}$ and $D_{max}^{c2}$ constraints are relaxed to allow more bounding-boxes to be added to the confirmed sets, \textbf{that is,} any matched pairs found in the EMD matching are used for adding the candidate bounding-boxes to the confirmed bounding-box sets. Specifically, for each confirmed $c1$ bounding-box associated with $q_j$, where $1 \leq j \leq t$, we find $p_{i^*}$ such that $f_{i^*,j} = 1$, and then add the corresponding candidate bounding-box $p_{i^*}$ to the confirmed $c1$ bounding-box set. The same procedure is applied for $c2$ bounding-boxes associated with $q_j$, where $t < j \leq m$.

The termination condition for multiple rounds of bounding-box propagation occurs when the number of candidate bounding-boxes in $P$ becomes 0. At this point, we report all the bounding boxes in the confirmed bounding-box sets for both $c1$ and $c2$ class categories to form the final detection results.

\vspace{-0.5mm}
\section{Experiments}
\label{sec:Experiments}
\subsection{Datasets}
To validate the effectiveness of our proposed method on histology images, we conduct experiments on two datasets: GlaS~\cite{gland} and MoNuSeg~\cite{cell}. The GlaS dataset consists of 165 images from 16 colon adenocarcinoma tissue slides. Among them, 85 images are used as the training set, 60 images are used as test set A, and 20 images are used as test set B. The MoNuSeg dataset contains different cell nuclei from multiple organs. It includes 30 images (around 22,000 cells) for training and 14 images (around 7000 cells) for testing. The codes for all the experiments will be made publicly available.

\vspace{-0.3cm}
\subsection{Main Results}
We use the bounding-box predictions from a well-trained DiffusionDet~\cite{diffusiondet} with a confidence threshold set at $0.50$ as the baseline. The inputs for the proposed TSBP method are obtained from the same DiffusionDet model with all of its output bounding-boxes (without applying the threshold). We compare TSBP with a range of confidence calibration methods, Logic Calibration (LC)~\cite{Prob,face}, Beta Calibration (BC)~\cite{beta,face}, and Histogram Binning (HB)~\cite{HB,face}\footnote{{https://github.com/EFS-OpenSource/calibration-framework}}. Since calibration methods require additional labeled data for parameter estimations, we allocate $20\%$ of the samples in the test sets for parameter estimations for calibration methods. We then perform model evaluation (including our method) using the remaining $80\%$ of samples in the test sets. For confidence calibration methods, after calibration, we use a threshold set at 0.50 to decide which bounding-boxes to include in their final detection outputs.

\begin{table}[!t]
    \centering
    \footnotesize
    \caption{F-scores ($\%$) of the object detection on the GlaS and MoNuSeg datasets.}
        \begin{tabular}{|ccccccc}
             \hline
             \multicolumn{2}{|c}{Dataset} & \multicolumn{1}{|c}{Baseline} & \multicolumn{1}{|c}{LC~\cite{Prob}  } & \multicolumn{1}{|c}{BC~\cite{beta}} & \multicolumn{1}{|c}{HB~\cite{HB}} & \multicolumn{1}{|c|}{TSBP}\\
             \hline
             \multicolumn{2}{|c}{Labeled Samples Used} & \multicolumn{1}{|c}{$0\%$} & \multicolumn{1}{|c}{$20\%$} & \multicolumn{1}{|c}{$20\%$} & \multicolumn{1}{|c}{$20\%$} & \multicolumn{1}{|c|}{$0\%$}\\
             \hline
             \multirow{2}{*}{GlaS} & \multicolumn{1}{|c}{testA} & \multicolumn{1}{|c}{85.94} & \multicolumn{1}{|c}{89.37} & \multicolumn{1}{|c}{87.76} & \multicolumn{1}{|c}{89.46} & \multicolumn{1}{|c|}{\textbf{90.39}}\\
             \cline{2-7}
             ~ & \multicolumn{1}{|c}{testB} & \multicolumn{1}{|c}{69.18} & \multicolumn{1}{|c}{72.63} & \multicolumn{1}{|c}{66.67} & \multicolumn{1}{|c}{68.35} & \multicolumn{1}{|c|}{\textbf{73.45}}\\
             \hline
             \multicolumn{2}{|c}{MoNuSeg} & \multicolumn{1}{|c}{83.23} & \multicolumn{1}{|c}{81.66} & \multicolumn{1}{|c}{82.85} & \multicolumn{1}{|c}{83.35} & \multicolumn{1}{|c|}{\textbf{83.60}}\\
             \hline
        \end{tabular}
    \label{table1}
\end{table}

\begin{figure*}[!tb]
    \centering
    \vspace{1mm}
    \includegraphics[width=\textwidth]{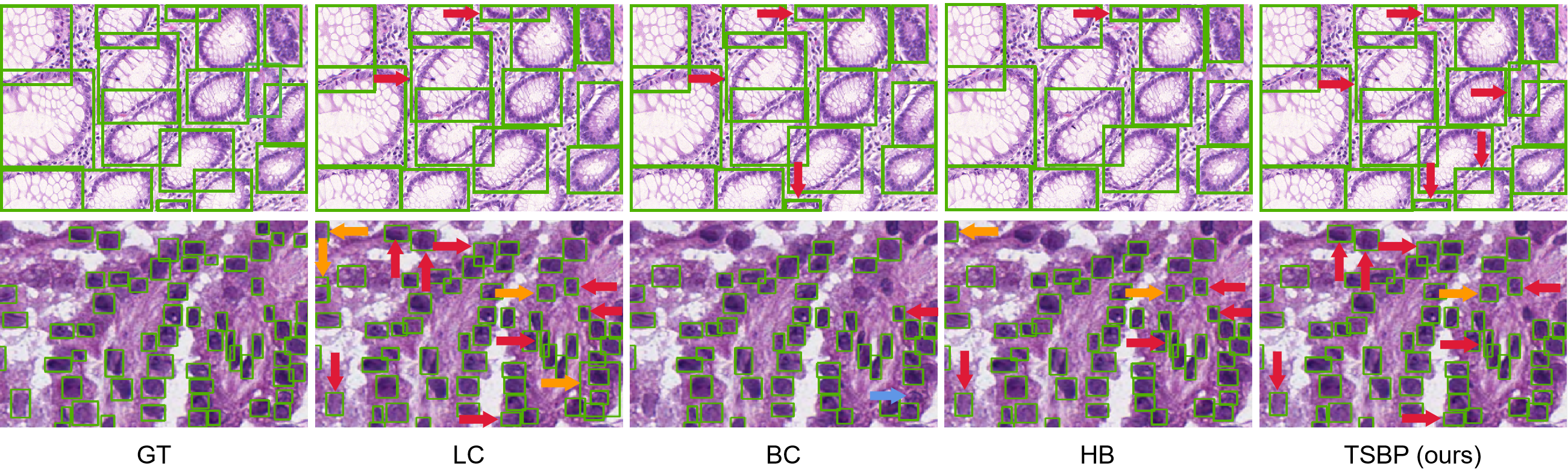}
    \caption{Top: Gland detection. Bottom: Cell nucleus detection. The red arrows indicate additional true positives compared to the baseline results. The orange arrows indicate additional false positives compared to the baseline results. The blue arrow indicates missed detections compared to the baseline results.}
    \vspace{-3mm}
    \label{fig3}
\end{figure*}

Table~\ref{table1} summarizes the quantitative results of these methods, it can be observed that TSBP achieves the highest F-scores on both the tested datasets. Note that the calibration methods require \textbf{additional labeled data} for parameter estimation, while TSBP does not use such additional labeled samples. 
Fig.~\ref{fig3} shows some visual results of TSBP and the competing methods. Compared to the baseline method, TSBP further increases the number of true positives. In the first row of the visual results, our method has the highest number of additional true positives, and TSBP finds all the remaining missed detection boxes compared to the other results. In the second row of the prediction images, the LC method has the highest number of additional true positives (LC adds eight true positives, while TSBP adds seven true positives), but, the LC method introduces more false positives (LC adds four false positives, while TSBP adds one false positives). In comparison, our method achieves a better balance between missed detections and false positives. 

\subsection{Ablation and Additional Studies}
\textbf{On Initial Bounding-Box Selection.} The parameter $s^{c1}$ and $s^{c2}$ determine the inclusion criteria for bounding-boxes in the initial confirmed bounding-box set. Here, we examine the impact of selecting the initial confidence threshold $s^{c1}$ on the detection performance. As shown in Table~\ref{table2}, TSBP yields better detection results with a lower initial threshold (0.60). When a higher threshold (0.70) is applied, the initial confirmed bounding-boxes may fail to encompass the diversity of object appearances in the test samples, resulting in sub-optimal performance of bounding-box propagation. Notably, compared to other methods across varying confidence thresholds, TSBP consistently achieves better results.

\begin{table}[!t]
    \centering
    \scriptsize
    \caption{F-scores ($\%$) with varying $s^{c1}$ used in TSBP. For comparison, other methods use $s^{c1}$ as the threshold for selecting bounding-boxes.}
        \begin{tabular}{|cccccccccc}
             \hline
             \multicolumn{2}{|c}{\multirow{2}{*}{~}} & \multicolumn{3}{|c}{ threshold= 0.70} & \multicolumn{1}{|c}{ $s^{c1}$= 0.70} & \multicolumn{3}{|c}{ threshold= 0.60} & \multicolumn{1}{|c|}{$s^{c1}$ = 0.60} \\
             \cline{3-10}
             ~ & ~ & \multicolumn{1}{|c} {Baseline} & \multicolumn{1}{|c}{LC~\cite{Prob}} &  \multicolumn{1}{|c}{HB~\cite{HB}} &\multicolumn{1}{|c}{TSBP} & \multicolumn{1}{|c}{Baseline} & \multicolumn{1}{|c}{LC~\cite{Prob}} &  \multicolumn{1}{|c}{HB~\cite{HB}}  & \multicolumn{1}{|c|}{TSBP} \\
             \hline
             \multicolumn{2}{|c}{Labeled Samples Used} & \multicolumn{1}{|c}{$0$} & \multicolumn{1}{|c}{$20\%$} & \multicolumn{1}{|c}{$20\%$} &  \multicolumn{1}{|c}{\textbf{$0$}} & \multicolumn{1}{|c}{\textbf{$0$}} & \multicolumn{1}{|c}{$20\%$} &  \multicolumn{1}{|c}{$20\%$}  & \multicolumn{1}{|c|}{$0$}  \\
             \hline
             \multirow{2}{*}{GlaS} & \multicolumn{1}{|c}{testA} & \multicolumn{1}{|c}{69.45} & \multicolumn{1}{|c}{82.34} &  \multicolumn{1}{|c}{86.97} & \multicolumn{1}{|c}{\textbf{88.42}} & \multicolumn{1}{|c}{79.30} &  \multicolumn{1}{|c}{86.78} &  \multicolumn{1}{|c}{86.97}&\multicolumn{1}{|c|}{\textbf{89.29}} \\
             \cline{2-10}
             ~ & \multicolumn{1}{|c}{testB} & \multicolumn{1}{|c}{51.13}  & \multicolumn{1}{|c}{68.35} &  \multicolumn{1}{|c}{65.36} & \multicolumn{1}{|c}{\textbf{72.73}} &\multicolumn{1}{|c}{59.31}  & \multicolumn{1}{|c}{70.66}  & \multicolumn{1}{|c}{68.35} & \multicolumn{1}{|c|}{\textbf{72.73}} \\
             \hline
             \multicolumn{2}{|c}{MoNuSeg} & \multicolumn{1}{|c}{72.62} & \multicolumn{1}{|c}{79.97} &  \multicolumn{1}{|c}{80.59} & \multicolumn{1}{|c}{\textbf{80.83}} & \multicolumn{1}{|c}{80.59}  & \multicolumn{1}{|c}{83.20}  &  \multicolumn{1}{|c}{83.35} & \multicolumn{1}{|c|}{\textbf{83.76}}  \\
             \hline
        \end{tabular}
    \label{table2}
\end{table}

\noindent \textbf{$K$ in the K-means step of TSBP.} In Table~\ref{table3}, we show that different values of $K$ lead to performance fluctuations in a small range on the GlaS (testA) and MoNuSeg datasets. Due to the smaller size of the testB in GlaS dataset, smaller value of $K$ gives better results.

\noindent \textbf{Error rates in different stages of TSBP.} Table~\ref{table4} demonstrates that the error rate in the first stage is lower than that in the second stage, consistent with the method design, where early propagated samples are more likely to be assigned correct labels.

\begin{table}[!t]
    \centering
    \footnotesize
    \caption{Resulting F-score (\%) with varying $K$ in the K-means step of TSBP.}
        \begin{tabular}{|cccccc}
             \hline

             \multirow{2}{*}{GlaS} & \multicolumn{1}{|c}{testA} & \multicolumn{1}{|c}{90.21 ($K=15$)} & \multicolumn{1}{|c}{90.11 ($K=20$)} & \multicolumn{1}{|c}{90.39 ($K=25$)} & \multicolumn{1}{|c|}{89.75 ($K=30$)} \\
             \cline{2-6}
             ~ & \multicolumn{1}{|c}{testB} & \multicolumn{1}{|c}{73.45 ($K=5$)} & \multicolumn{1}{|c}{73.14 ($K=10$)} & \multicolumn{1}{|c}{72.94 ($K=20$)} & \multicolumn{1}{|c|}{70.93 ($K=25$)} \\
             \hline
             \multicolumn{2}{|c}{MoNuSeg} & \multicolumn{1}{|c}{83.52 ($K=15$)} & \multicolumn{1}{|c}{83.56 ($K=20$)} & \multicolumn{1}{|c}{83.60 ($K=25$)} & \multicolumn{1}{|c|}{83.71 ($K=30$)}\\
             \hline
        \end{tabular}
    \label{table3}
\end{table}

\begin{table}[!t]
    \centering
    \footnotesize
    \caption{Error rates ($\%$) of detection in different stages of TSBP.}
        \begin{tabular}{|cccc}
             \hline
             \multicolumn{2}{|c}{GlaS} & \multicolumn{2}{|c|}{MoNuSeg}  \\
             \hline
              \multicolumn{1}{|c}{26.79  (first satge)} & \multicolumn{1}{|c}{37.25 (second stage)} & \multicolumn{1}{|c}{50.66 (first stage)} & \multicolumn{1}{|c|}{53.44 (second stage)}  \\
             \hline
        \end{tabular}
    \vspace{-2mm}
    \label{table4}
\end{table}

\vspace{-0.1cm}
\section{Conclusion}
\vspace{-0.1cm}
The abundance of visual information in histology images can be leveraged during test time to improve the performance of key object detection. This paper introduced a novel perspective and approach to enhance object detection in histology images in test time. Specifically, we designed a versatile matching-based algorithm (TSBP) that leverages detections with high model confidence to influence detections with low model confidence, aiming to rectify potential errors and enhance the overall accuracy of the detection results. Experiments on two object detection tasks in histology images showed the effectiveness of our proposed method. Compared to confidence calibration methods, our method requires no additional labeled data for model training and parameter estimation. \textcolor{black}{Future work will focus on developing a more automatic and robust procedure for forming the initial condition of TSBP.} Further studies of utilizing TSBP for test-time model training can be considered an interesting future research target.

\bibliographystyle{splncs04}
\bibliography{refs}
%




\end{document}